\documentclass{emulateapj}
\usepackage{epsfig}
\usepackage[figuresright]{rotating}
\usepackage{natbib}
\usepackage{xcolor}
\usepackage[colorlinks, citecolor=blue]{hyperref}
\usepackage{longtable}
\usepackage{url}
\usepackage{lineno}
\usepackage{amsmath}

\shorttitle{Kilonova emission from compact star mergers}
\shortauthors{Wang et al.}

\begin{document}

\title{Ejecta$-$circumstellar medium interaction in high-density environment contribution to kilonova
emission:
Application to GRB 191019A}
\author{Suo-Ning Wang\altaffilmark{1}, Hou-Jun L\"{u}\altaffilmark{1}, Yong Yuan\altaffilmark{2}, Hao-Yu
Yuan\altaffilmark{3}, Jared Rice\altaffilmark{4}, Meng-Hua Chen\altaffilmark{1}, and En-Wei
Liang\altaffilmark{1}}

\altaffiltext{1}{Guangxi Key Laboratory for Relativistic Astrophysics, School of Physical Science
and Technology, Guangxi University, Nanning 530004, China; lhj@gxu.edu.edu} \altaffiltext{2}{School
of Physics Science And Technology, Wuhan University, No.299 Bayi Road, 430072, Wuhan, Hubei, China}
\altaffiltext{3}{School of Physics, Huazhong University of Science and Technology, Wuhan 430074,
China}\altaffiltext{4}{Department of Mathematics and Physical Science, Southwestern Adventist
University, Keene, TX 76059, USA}

\begin{abstract}	
The nearby long-duration GRB 191019A recently detected by Swift lacks an associated supernova and
belongs to a host galaxy with little star formation activity, suggesting that the origin of this
burst is the result of a merger of two compact objects with dynamical interactions in a
high-density medium of an active galactic nucleus. Given the potential motivation of this event,
and given that it occurs in such a high-density environment, the ejecta-circumstellar medium
(CSM) interaction cannot be ignored as possibly contributing to the kilonova emission. Here, we
theoretically calculate the kilonova emission by considering the contribution of the ejecta-CSM
interaction in a high-density environment. We find that the contribution to the kilonova emission
from the ejecta-CSM interaction will dominate at a later time, and a smaller ejecta mass will
have a stronger kilonova emission from the ejecta-CSM interaction. Moreover, we try to apply it
to GRB 191019A, but we find that it is difficult to identify the possible kilonova emission from
the observations, due to the contribution of the bright host galaxy. On the other hand, less
injected mass (less than $M_{\rm ej}=2\times10^{-5}M_{\odot}$) will be required if one can detect
the kilonova emission associated with a GRB 191019A-like event in the future. The {\em
r}-process-powered and spin energy contributions from the magnetar are also discussed.

\end{abstract}
\section{Introduction}

Merger of two compact stars, such as neutron star$-$neutron star (NS-NS) mergers
\citep{1986ApJ...308L..43P,1989Natur.340..126E} or neutron star$-$black hole (NS-BH) mergers
\citep{1991AcA....41..257P}, are thought to be potential sources of gravitational-wave
(GW) radiation, short-duration gamma-ray bursts (GRBs), as well as optical/infrared transients
(called kilonovae; \citealt{2014ARA&A..52...43B}; \citealt{2018pgrb.book.....Z}). The first
``smoking gun" evidence to support this hypothesis was the direct detection of a GW event
(GW170817) by Advanced LIGO and Virgo that was associated with the short GRB 170817A and kilonova
AT2017gfo
\citep{2017PhRvL.119p1101A,2017NatAs...1..791C,2017Sci...358.1565E,2017ApJ...848L..14G,2017Natur.551...80K,2017ApJ...848L..15S,2018NatCo...9..447Z}.

Phenomenally, GRBs are divided into those of long and
short durations, with a division line at the observed duration $T_{90}\sim$2 s \citep{1993ApJ...413L.101K},
and it has been suggested that they are likely
related to compact star merger and the deaths of massive stars, respectively
\citep{2018pgrb.book.....Z}. However, the duration of a GRB is energy- and instrument-dependent
\citep{2013ApJ...763...15Q}, suggesting that it may not be directly related to the
progenitors of GRBs \citep{2010ApJ...725.1965L,2014MNRAS.442.1922L}. It requires a multiple
observational criteria to judge the progenitors of GRBs \citep{2009ApJ...703.1696Z}. A small
fraction of short-duration GRBs were reported to be associated with kilonova candidates before the
discovery of GW 170817/GRB 170817A/AT2017gfo (see \citealt{2019LRR....23....1M} for a review)$-$for
example, GRB 130603B \citep{2013ApJ...774L..23B,2013Natur.500..547T}, GRBs 050709 and 070809
\citep{2016NatCo...712898J,2020NatAs...4...77J}, GRB 080503
\citep{2014MNRAS.439.3916M,2015ApJ...807..163G}, and GRB 160821B
\citep{2017ApJ...835..181L,2019ApJ...883...48L,2019MNRAS.489.2104T}. On the other hand, excepting the
above short-duration GRBs, several nearby short-duration GRBs with extended emission (EE) lacking
an associated supernova but containing a possible kilonova signature were claimed to be from
compact star mergers$-$for example, GRB 060614
\citep{2006Natur.444.1050D,2006Natur.444.1047F,2006Natur.444.1053G,2006Natur.444.1044G,2015NatCo...6.7323Y},
GRB 211211A
\citep{2022Natur.612..223R,2022Natur.612..228T,2022Natur.612..232Y,2023ApJ...943..146C,2023NatAs...7...67G},
and GRB 211227A \citep{2022ApJ...931L..23L,2023A&A...678A.142F}.

Recently, a nearby peculiar long-duration GRB 191019A without EE emission that triggered the Swift
Burst Alert Telescope (BAT; \citealt{2019GCN.26031....1S}) with redshift $z=$0.248
\citep{2023NatAs...7..976L} has generated excitement. More interestingly, no associated supernova
signature was detected for GRB 191019A and the host galaxy has no significant star
formation activity \citep{2023NatAs...7..976L}. This suggests that the origin of GRB 191019A is
unlikely to be the death of a massive star \citep{2023NatAs...7..976L}, but that dynamical
interactions in the dense nucleus of the host via compact object mergers could naturally explain
the observations \citep{2023NatAs...7..976L}. For example, gas drag \citep{2020ApJ...898...25T}
and the clustering of compact objects in migration traps promote binary formation through
dynamic interactions \citep{2016ApJ...819L..17B,2020MNRAS.498.4088M}. \cite{2023ApJ...950L..20L}
proposed that the GRB 191019A originated in a binary compact merger with an intrinsic prompt
emission duration of $\sim 1.1$ s, which is stretched in time by the interaction with a
high-density external medium $\sim 10^7-10^8~\rm cm^{-3}$ \citep{2022ApJ...938L..18L}. If
this is the case, a kilonova emission associated with GRB 191019A should be powered after the
merger of the compact stars, and the contribution of an ejecta$-$circumstellar medium (CSM)
interaction producing a kilonova cannot be ignored in such a dense external medium
\citep{2022ApJ...925...43Q,2022ApJ...940L..44R}. In previous studies, the ejecta$-$CSM interaction
has been invoked to interpret the observations of some superluminous supernovae
\citep{1994ApJ...420..268C,2007ApJ...671L..17S,2011ApJ...729L...6C,2012ApJ...746..121C,2013ApJ...773...76C,2014MNRAS.444.2096N,2016ApJ...828...87W,2018ApJ...856...59L},
but never has never been used to interpret any kilonova emission associated with short
GRBs.

From a theoretical point of view, the kilonova emission from a merger of two compact stars is
generated by the ejected materials, powered by radioactive decay from the $r$-process, and is
nearly isotropic
\citep{1998ApJ...507L..59L,2010MNRAS.406.2650M,2011NewAR..55....1B,2011ApJ...732L...6R,2013PhRvD..87b4001H,2013MNRAS.430.2585R,2015ApJ...811L..22J}.
On the other hand, the magnetar central engine can be formed as the remnant, depending on the mass
of the nascent NS, if the merger of the two compact stars is NS-NS or an NS and a white dwarf
(NS-WD)
\citep{1992Natur.357..472U,1998PhRvL..81.4301D,2001ApJ...552L..35Z,2006Sci...311.1127D,2010MNRAS.409..531R}.
If this is the case, the spin energy of the magnetar can be comparable to or even larger than the
radioactive decay energy and may power a brighter kilonova (called a ``merger-nova'';
\citealt{2013ApJ...776L..40Y,2013ApJ...763L..22Z,2014MNRAS.439.3916M,2017ApJ...837...50G,2017ApJ...835..181L}),
such as 160821B \citep{2017ApJ...835..181L} and GRB 170817A \citep{2018ApJ...861..114Y}. It is also
possible that the magnetic wind from the accretion disk of a newborn BH could heat up the
neutron-rich merger ejecta and provide additional energy to the transients if the central engine is
a BH for NS-BH or NS-NS mergers
\citep{2015NatCo...6.7323Y,2021ApJ...911...97M,2021ApJ...912...14Y}, such as GRB 160821B
\citep{2021ApJ...911...97M,2021ApJ...912...14Y}.

If the merger of two compact stars is indeed operating in GRB 191019A, then a question arises: how
bright is the kilonova emission associated with GRB 191019A when we simultaneously consider the
different types of energy sources, i.e. $r$-process-powered, spin energy from a magnetar, and the
contribution from an ejecta$-$CSM interaction? This paper aims to address this interesting question
through detailed calculations of the kilonova emission, which are then compared with the calculated
results for the kilonova from the observational data of GRB 191019A. This paper is organized as
follows. The models of the kilonova energy sources we use are shown in Section \ref{section 2}, and
Section \ref{section 3} presents the basic observations of GRB 191019A. In Section \ref{section 4},
we show the results of the kilonova calculations to compare with the observational data of GRB
191019A. Conclusions are drawn in Section \ref{section 5} with some additional discussions.
Throughout the paper, we use the notation $Q_{n}=Q/10^{n}$ in CGS units, which means that $Q_{n}$
is the unit of $10^{n}$ (such as $E_{52}=E/(10^{52}~\rm erg)$). We adopt a concordance cosmology
with parameters $H_{\rm 0}=71~\rm km~s^{-1}~Mpc^{-1}$, $\Omega_{\rm M}=0.30$, and
$\Omega_{\Lambda}=0.70$.

\section{Basic model of kilonova energy sources}
\label{section 2}
\subsection{$\rm r$-process nucleosynthesis}
The ejecta from NS-NS or NS-BH mergers carries a large amount of free neutrons. The ejecta expands
in space at a nearly constant speed and the neutron capture time scale is shorter than the
$\beta$-decay time scale in such a dense neutron-rich environment. This allows the rapid neutron
capture process (called $r$-process) to quickly synthesize elements heavier than Fe. The heavier
elements are unstable and heat the ejecta through radioactive decay. Photons are trapped in the
ejecta, but are eventually released from the photosphere to power the kilonova, which radiates in
the ultraviolet, optical, and infrared bands
\citep{1998ApJ...507L..59L,2010MNRAS.406.2650M,2017LRR....20....3M}.

The total energy of the ejecta can be written as
$E_{\mathrm{ej}}=(\Gamma-1)M_{\mathrm{ej}}c^{2}+\Gamma E_{\mathrm{int}}^{\prime}$, where
$E_{\mathrm{int}}^{\prime}$ is the internal energy measured in the comoving frame,
$M_{\mathrm{ej}}$ is the ejecta mass, $c$ is the speed of light, and $\Gamma$ is the Lorentz
factor. Based on energy conservation, one has
\begin{eqnarray}
    dE_{\mathrm{ej}}=(L_{\mathrm{inj}}-L_{\mathrm{e}})dt,
\end{eqnarray}
where $L_{\rm e}$ and $L_{\mathrm{inj}}$ are the radiated bolometric luminosity and the injection
luminosity, respectively. Within the $r$-process scenario, the injection luminosity is contributed to by
radioactive power ($L_{\mathrm{ra}}$):
\begin{eqnarray}
    L_\text{inj}=L_{\mathrm{ra}}.
\end{eqnarray}
The radioactive power in the comoving frame can be defined as
\begin{eqnarray}
\begin{split}
L_{\text{ra}}' &= \frac{L_{\text{ra}}}{\mathcal{D}^2}\\&=4\times10^{49}M_{\text{ej},-2}
\times\left[\frac{1}{2}-\frac{1}{\pi}\arctan\left(\frac{t'-t_0'}{t_\sigma'}\right)\right]^{1.3}\text{erg
s}^{-1},
\end{split}
\end{eqnarray}
where $t_0'\sim$1.3 s and $t_\sigma'\sim0.11$ \citep{2012MNRAS.426.1940K}, and $\mathcal{D} =
1/[\Gamma(1-\beta)]$ is the Doppler factor with $\beta=\sqrt{1-\Gamma^{-2}}$. The comoving time
$dt'$ can be connected with the observer's time via $d t^{\prime}={\cal D}d t$.

Together with the above considerations, the equation for the dynamic evolution of the ejecta can be
expressed as
\begin{eqnarray}
    \frac{d\Gamma}{dt}=\frac{{L_{\mathrm{inj}}}-L_{\rm
    e}-\Gamma\mathcal{D}(dE_{\mathrm{int}}'/dt')}{M_{\mathrm{ej}}c^2+E_{\mathrm{int}}'}.
\end{eqnarray}
The evolution of internal energy in the comoving frame can be written as
\citep{2010ApJ...717..245K, 2013ApJ...776L..40Y}
\begin{eqnarray}
    \frac{dE'_{\mathrm{int}}}{dt'}=L'_{\mathrm{inj}}-L'_{\mathrm{e}}-\mathcal{P'}\frac{dV'}{dt'}.
\end{eqnarray}
The pressure $\mathcal{P}'=E_\mathrm{int}'/3V'$ is dominated by radiation, and the evolution of the comoving
volume can be written as
\begin{eqnarray}
    \frac{dV'}{dt}=4\pi R^2\beta c,
\end{eqnarray}
where $R$ is the radius of the ejecta and $\beta$ is the velocity of the kilonova ejecta, which is equal
to $v_{\rm KN}$. One has
\begin{eqnarray}
    \frac{dR}{dt}=\frac{\beta c}{(1-\beta)}
\end{eqnarray}
such that the radiated bolometric luminosity in the comoving frame can be expressed as
\citep{2010ApJ...717..245K,2013MNRAS.432.3228K}
\begin{eqnarray}
    L_\mathrm{e}'=\begin{cases}\frac{E_\mathrm{int}'c}{\tau R/\Gamma},&t\leqslant t_\tau\\
    \frac{E_\mathrm{int}'c}{R/\Gamma},&t>t_\tau\end{cases},
\end{eqnarray}
where $\tau = \kappa(M_{\mathrm{ej}}/V')(R/\Gamma)$ is the optical depth of the ejecta, $\kappa$ is
the opacity, and $t_\tau$ is the time at ${\tau}=1$.

The radiation of the ejecta comes from the photosphere $R_{\rm ph}$ with a blackbody radiation
spectrum. The effective temperature can be defined as \citep{2018ApJ...861..114Y}
\begin{eqnarray}
    T_{\rm e}=\left(\frac{L_{\rm e}}{4\pi R_\text{ph}^2\sigma_\text{SB}}\right)^{1/4},
\end{eqnarray}
where $\sigma_\text{SB}$ is the Stephan$-$Boltzman constant. The observed flux density at frequency
${\nu}$ can be calculated as
\begin{eqnarray}
    F_\nu^{\rm ejcta}=\frac{2\pi h\nu^3R^2}{D_L^2c^2}\frac{1}{\exp(h\nu/kT_{\rm e})-1},
\end{eqnarray}
where $h$ is the Planck constant and $D_L$ is the luminosity distance of the source.

\subsection{Spindown energy}
For a stable NS with an initial spin period $P_{\rm i}$, the total rotational energy is
$E_{\mathrm{rot}}\sim2\times10^{52}I_{45}M_{1.4}R_{6}^{2}P_{\mathrm{i.-3}}^{-2} ~\rm erg$, where
$M$, $R$, and $I$ are the mass, radius, and moment of inertia of the NS, respectively.
The spindown luminosity of a stable NS as a function of time can be expressed by the magnetic
dipole radiation formula equation \citep{2001ApJ...552L..35Z,2018MNRAS.480.4402L}:
\begin{eqnarray}
    L_{\mathrm{sd}}=L_0\bigg(1+\frac{t}{t_{\rm sd}}\bigg)^\alpha
\end{eqnarray}
where $L_0\simeq1.0\times10^{47}~\operatorname{erg~s}^{-1}(B_{p,14}^2P_{i,-3}^{-4}R_6^6)$, and
$t_{\rm sd}\simeq2\times10^5~\mathrm{s}(I_{45}B_{p,14}^{-2}P_{0,-3}^2R_6^{-6})$ are the
characteristic spindown luminosity and timescale, respectively. $t$ is the time in the observed
frame and $\alpha$ is equal to $-1$ and $-2$ when the energy loss is dominated by GW radiation
\citep{2016MNRAS.458.1660L,2020ApJ...898L...6L} and magnetic dipole radiation
\citep{2001ApJ...552L..35Z}, respectively. Sometimes, $\alpha$ may be less than $-2$, which
corresponds to the situation of a supermassive NS central engine collapsing into a BH
\citep{2010MNRAS.409..531R,2021ApJ...912...14Y}.

If the central engine of a short GRB is a stable NS, the main power source of the kilonova should
no longer be limited to the $r$-process and the spin energy of the NS may be comparable to or even
larger than the radioactive decay energy. In this case, the ejection luminosity should have
contributions from both the spindown luminosity ($L_{\mathrm{sd}}$) and radioactive power
($L_{\mathrm{ra}}$):
\begin{eqnarray}
    L_\text{inj}={\xi}L_{\mathrm{sd}}+L_{\mathrm{ra}},
\end{eqnarray}
where ${\xi}$ corresponds to the converted efficiencies (fixed to 0.3) from spindown luminosity to
radiation \citep{2011ApJ...726...90Z}. The subsequent operations are the same as in the previous
part.

\subsection{Ejecta$-$CSM interaction}
If the external medium of a short GRB is dense enough, we have to consider the contribution of an
ejecta$-$CSM interaction to the kilonova emission. The ejecta$-$CSM interaction
model involving multiple shells/winds has been invoked to interpret supernova light curves with
multiple peaks \citep{2016ApJ...828...87W,2018ApJ...856...59L}. Following the same method of
ejecta$-$CSM interaction that has been used in supernova explosions, we invoke the ejecta$-$CSM
interaction model involving one shell to calculate the kilonova emission. In this section, we will
present the details of this ejecta$-$CSM interaction model.

The ejecta interacting with the pre-existing CSM can result in two shocks: a forward shock (FS)
that propagates through the CSM and a reverse shock (RS) that sweeps up the kilonova ejecta. Based
on numerical simulations of supernova explosions, the density of the supernova ejecta is a
power-law distribution \citep{1982ApJ...258..790C,1994ApJ...420..268C,2018ApJ...856...59L}. We
assume that the density distribution of the kilonova ejecta also obeys a power-law distribution
\citep{1999ApJ...510..379M}. Thus, the density profile of the outer part of the kilonova ejecta is
written as
\begin{eqnarray}
    \rho_{\mathrm{out}}=g_{\rm n} t^{n-3}r^{-n},
\end{eqnarray}
where $n$ is the power-law index of the outer part of the ejecta that depends on the kilonova
progenitor star. $g_{\rm n}$ is the scaling parameter of the density profile, and it is given by
\citep{1994ApJ...420..268C,2012ApJ...746..121C}
\begin{eqnarray}
    g_{\rm
    n}=\frac{1}{4\pi(n-\delta)}\frac{[2(5-\delta)(n-5)E_{\mathrm{KN}}]^{(n-3)/2}}{[(3-\delta)(n-3)M_{\mathrm{ej}}]^{(n-5)/2}},
\end{eqnarray}
where $\delta$ is the power-law index of the inner density profile and $E_{\mathrm{KN}}$ is the
total kilonova energy. The density of a circumstellar shell is written as
\begin{eqnarray}
    \rho_{\mathrm{CSM}}=qr^{-s},
\end{eqnarray}
where $q$ is a scaling constant and $s$ is the power-law index for a CSM density profile. $s=0$
represents a uniform-density shell. Here, we adopt a uniform-density shell ($s=0$) to do the
calculations and obtain the density of the ambient material $\rho_{\mathrm{CSM}}=q$. Moreover, in
order to test how sensitive the kilonova emission is for $n$, we calculate the kilonova emission
that is contributed from the ejecta$-$CSM interaction with varying $n$. We find that a smaller-
quality $n$ is corresponding to a longer time and brighter emission from the CSM interaction. The
kilonova emission from the ejecta$-$CSM interaction with $n=7$ is comparable with observations, so we
adopt $n=7$ to do the calculations.

A contact discontinuity separates the shocked CSM and the shocked ejecta. Its radius ($R_{\rm cd}$)
can be described by the self-similar solution \citep{1982ApJ...258..790C}
\begin{eqnarray}
    R_{\mathrm{cd}}=\left(\frac{A g_{\rm n}}{q}\right)^{\frac{1}{n-s}}t^{\frac{(n-3)}{(n-s)}},
\end{eqnarray}
where $A$ is a constant. The radii of the FS and RS are written as
\begin{eqnarray}
    R_{\mathrm{FS}}(t)=R_{\mathrm{in}}+\beta_{\mathrm{FS}}R_{\mathrm{cd}}
\end{eqnarray}
and
\begin{eqnarray}
    R_{\mathrm{RS}}(t)=R_{\mathrm{in}}+\beta_{\mathrm{RS}}R_{\mathrm{cd}},
\end{eqnarray}
where $R_{\rm in}$ is the initial radius of the interaction. $\beta_{\mathrm{FS}}$ and
$\beta_{\mathrm{RS}}$ are constants representing the ratio of the shock radius to the contact
discontinuity radius, respectively. The values of $\beta_{\mathrm{FS}}$ and $\beta_{\mathrm{RS}}$
are shown in Table 1 of \cite{1982ApJ...258..790C}. The self-similar solutions for the structure of
the interaction region require to $n>5$. Based on the result in \cite{2018ApJ...856...59L}, we
adopt $n=7$, which is used in the outer part of type I supernovae. Given $n=7$ and $s=0$, one has
$\beta_{\mathrm{FS}}=1.181$, $\beta_{\mathrm{RS}}=0.935$, and $A=1.2$. The ejecta$-$CSM interaction
will convert the kinetic energy to radiation. The input luminosity function of the FS is
\citep{2012ApJ...746..121C}
\begin{eqnarray}
    \begin{split}
        L_{\mathrm{FS}}(t)=\frac{2\pi}{(n-s)^3}g_{\rm
        n}^{\frac{5-s}{n-s}}q^{\frac{n-5}{n-s}}(n-3)^2(n-5)\\\beta_{\mathrm{FS}}^{5-s}A^{\frac{n-5}{n-s}}\times(t+t_{\mathrm{int}})^{\alpha_i}\theta(t_{\mathrm{FS},\mathrm{BO}}-t),
    \end{split}
\end{eqnarray}
and for the RS it is \citep{2016ApJ...828...87W}
\begin{eqnarray}
    \begin{split}
        L_{\mathrm{RS}}(t)=2\pi\left(\frac{Ag_{\rm
        n}}{q}\right)^{\frac{5-n}{n-s}}\beta_{\mathrm{RS}}^{\mathrm{5-n}}g_{\rm
        n}\left(\frac{n-5}{n-3}\right)\\
        \times\left(\frac{3-s}{n-s}\right)^3(t+t_{\mathrm{int}})^{\alpha_i}\theta(t_{\mathrm{RS},\star}-t),
    \end{split}
\end{eqnarray}
where $\theta(t_{\mathrm{RS},*}-t)$ and $\theta(t_{\mathrm{FS},\mathrm{BO}}-t)$ represent the
Heaviside step function controlling the end times of the FS and RS, respectively.
$t_{\mathrm{int}}\approx R_{\mathrm{in}}/v_{\mathrm{SN}}$ is the time when the ejecta just touches
the CSM, representing the beginning of the ejecta$-$CSM interaction. The temporal index is
$\alpha_i=(2n+6s-ns-15)/(n-s)$. Here, we fix $n=7$ and $s=0$, such that one has $\alpha_{i} = -0.143$.

The RS termination timescale $t_{\mathrm{RS},\star}$ is defined as the time for the RS to sweep up
all of the effective ejecta \citep{2012ApJ...746..121C,2013ApJ...773...76C}:
\begin{eqnarray}
    t_{\mathrm{RS,}\ast}=\left[\frac{v_{\mathrm{KN}}}{\beta_{\mathrm{RS}}(Ag_n/q)^{\frac{1}{n-s}}}\left(1-\frac{(3-n)M_\mathrm{ej}}{4\pi
    v_{\mathrm{KN}}^{3-n}g_n}\right)^{\frac{1}{3-n}}\right]^{\frac{n-s}{s-3}}.
\end{eqnarray}
where $v_{\mathrm{KN}}$ is the velocity of the kilonova ejecta. This depends on both the ejecta mass and
the energy of the kilonova. Initially, we set $v_{\mathrm{KN}}$ as in the range of $0.1 c$ to $0.9 c$, and
find
that the apparent magnitude of the kilonova is affected a little bit by altering $v_{\mathrm{KN}}$. So, we
fixed vKN as $0.1 c$. The termination timescale of the FS is approximately equal to the time
of the FS breakout $t_{\mathrm{FS,BO}}$ when the optically thin\footnote{Here, the “optically thin”
means
that the optical depth is in the range of $2/3<\tau_{\rm CSM}<1$.} part of the CSM is swept up, and
is given as
 \citep{2012ApJ...746..121C,2013ApJ...773...76C}
\begin{eqnarray}
    \begin{aligned}
        t_{\mathrm{FS,BO}}
        =\{\frac{(3-s)q^{(3-n)/(n-s)}[A g_{\rm
        n}]^{(s-3)/(n-s)}}{4\pi\beta_{\mathrm{FS}}^{3-s}}\}^{\frac{n-s}{(n-3)(3-s)}}
         \times M_{\text{CSM},\text{th}}^{\frac{n-s}{(n-3)(3-s)}},
    \end{aligned}
 \end{eqnarray}
where $M_{\text{CSM},\text{th}}$ is the mass of the optically thin CSM:
\begin{eqnarray}
M_{\mathrm{CSM},\mathrm{th}}=\int_{R_{\mathrm{in}}}^{R_{\mathrm{ph}}}4\pi{r}^2\rho_{\mathrm{CSM}}dr.
\end{eqnarray}
Here, we fix $R_{\mathrm{in}}=10^{15}~\rm cm$. $R_{\mathrm{ph}}$ represents the photosphere radius
of the CSM shell. Under the Eddington approximation, it is located at an optical depth $\tau_{\rm
CSM}=2/3$ \citep{2018ApJ...856...59L}. $R_{\mathrm{ph}}$ is given by solving
\begin{eqnarray}
\begin{aligned}
\tau_{\rm CSM}=\int_{R_{\mathrm{ph}}}^{R_{\mathrm{out}}}\kappa_{\rm CSM}\rho_{\mathrm{CSM}}dr=2/3,
\end{aligned}
\end{eqnarray}
where $\kappa_{\mathrm{CSM}}$ is the optical opacity of the CSM and $R_{\operatorname{out}}$ is
the radius of the outer boundary of the CSM. The opacities of the CSM shells and winds
$\kappa_{\mathrm{CSM}}$ are related to their composition and temperatures. For hydrogen-poor
matter, the dominant source of opacity is electron scattering, $\kappa_{\mathrm{CSM}}=0.06-0.2~\rm
cm^{2}~g^{-1}$ (see the references listed in \citealt{2016ApJ...828...87W}). For hydrogen-rich
matter, $\kappa_{\mathrm{CSM}}=0.33~\rm cm^{2}~g^{-1}$, which is the Thomson electron scattering
opacity for fully ionized material with the solar metallicity \citep{2012ApJ...746..121C}. It is
given by solving
\begin{eqnarray}
    M_{\mathrm{CSM}}=\int_{R_{\mathrm{in}}}^{R_{\mathrm{out}}}4\pi r^2\rho_{\mathrm{CSM}}dr.
\end{eqnarray}\par

The total optical depth of the CSM depends on the opacity $\kappa_{\mathrm{CSM}}$ instead of
$M_{\mathrm{CSM}}$. For a given $M_{\mathrm{CSM}}=10M_{\odot}$ and hydrogen-rich matter
$\kappa_{\mathrm{CSM}}=0.33~\rm cm^{2}~g^{-1}$, the total optical depth of the
$\tau_{\mathrm{CSM}}$ is always larger than 1. If this is the case, one can only observe the
kilonova that is contributed from both the RS and FS of the ejecta$-$CSM
interaction. The interacting material could be heated by both the FS and RS, and contribute to the
input total luminosity. One has
\begin{eqnarray}
    L_{\mathrm{inp,CSM}}(t)=\epsilon[L_{\mathrm{FS}}(t)+L_{\mathrm{RS}}(t)],
\end{eqnarray}
where $\epsilon$ is the conversion efficiency from kinetic energy to radiation and is fixed at
$\epsilon=0.9$ to do the calculations, due to poor understanding. On the other hand, the optical
depth is very important for the kilonova emission. If $\tau_{\rm CSM}<2/3$, one can observe the
kilonova emission that is contributed from the $r$-process, spindown, as well as the RS of the
CSM, but not the contribution from the FS of the CSM. If $2/3<\tau_{\rm CSM}<1$, one can
observe the kilonova emission that is contributed from $r$-process, spindown, as well as both the
RS and FS of the ejecta$-$CSM interaction. If $\tau_{\rm CSM}>1$, one can
always observe the kilonova emission that is only contributed from the RS and FS of the
CSM, but one cannot observe the contributions from $r$-process and spindown. In order to observe the
kilonova emission that is contributed from $r$-process, spindown, as well as both the RS and
FS of the CSM, we set the total optical depth of the CSM $\tau_\textup{CSM}$ in the range of
$2/3<\tau_\textup{CSM}<1$ for $\kappa_{\mathrm{CSM}}=0.2~\rm cm^{2}~g^{-1}$. So that
$M_\textup{CSM,th}$ represents the mass of the optical depth larger than 2/3. If this is the case,
the output luminosity can be written as
\begin{eqnarray}
    L(t)=\frac{1}{t_{\rm diff}}\exp\left[-\frac{t}{t_{\rm diff}}\right]\int_0^t\exp\left[\frac{t'}{t_{\rm
    diff}}\right]L_{\mathrm{inp,CSM}}(t')dt',
\end{eqnarray}
where $t_{\rm diff}$ is the diffusion timescale. The interaction
diffusion timescale can be written as
\begin{eqnarray}
    t_{\mathrm{diff}}=\frac{\kappa_{\rm CSM} M_{\mathrm{CSM},\mathrm{th}}}{\phi cR_{\mathrm{ph}}},
\end{eqnarray}
where $\phi=4\pi^3/9\simeq13.8$ \citep{1982ApJ...253..785A}.\par

Based on the \equationautorefname{(11)} together with \equationautorefname{(12)}, one has
\begin{eqnarray}
    T_{\rm CSM}=\left(\frac{L(t)}{4\pi R_{\text{ph}}^2\sigma_{\text{SB}}}\right)^{1/4}
\end{eqnarray}
and
\begin{eqnarray}
    F_\nu^{\rm CSM}=\frac{2\pi h\nu^3R_{\rm ph}^2}{D_{\rm L}^2c^2}\frac{1}{\exp(h\nu/kT_{\rm CSM})-1}.
\end{eqnarray}
The observed kilonova emission flux consists of $F_\nu^{\text{ejcta}}$ which contains contributions
from the $r$-process nucleosynthesis, spindown energy, and $F_\nu^{\text{CSM}}$ from the
ejecta$-$CSM interaction:
\begin{eqnarray}
    F_\nu^{\text{tot}}=F_\nu^{\text{ejcta}}+F_\nu^{\text{CSM}}.
\end{eqnarray}
One may then determine the monochromatic apparent magnitude of the kilonova emission
\citep{2018ApJ...861..114Y}:
\begin{eqnarray}
    M_\nu=-2.5\log_{10}\frac{F_\nu^{\rm tot}}{3631}\text{Jy}.
\end{eqnarray}

\begin{figure}
\centering
 \includegraphics [angle=0,scale=0.5] {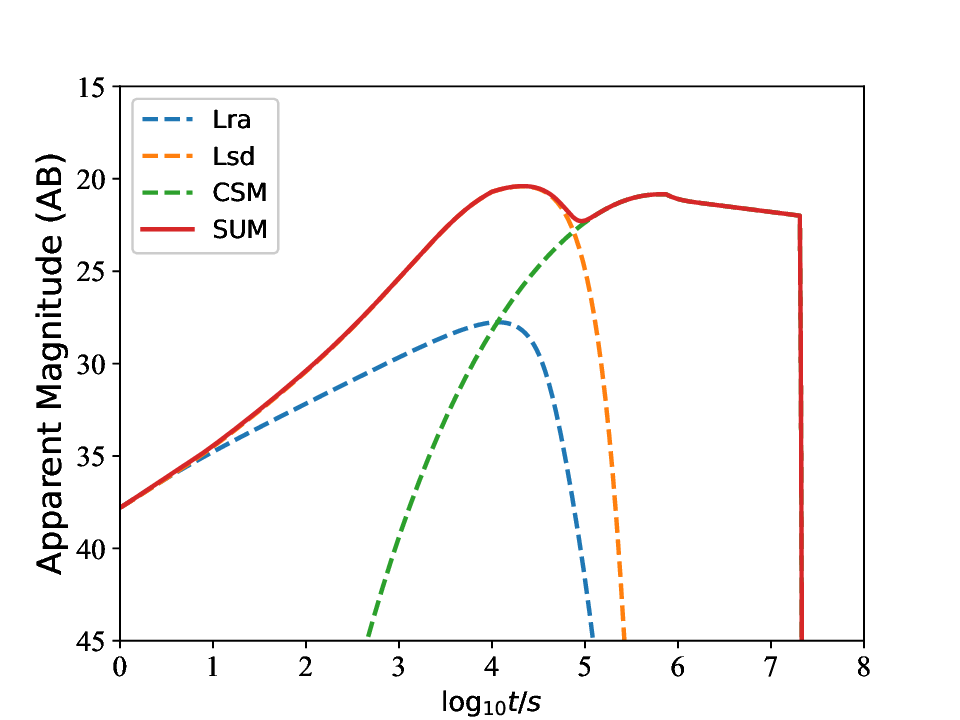}
 \caption{Kilonova $r$-band light curve considering the contributions from the three
 energy sources, e.g., the radioactive decay of heavier elements (the blue dashed line marked as $L_{\rm
 ra}$),
 the magnetar spindown energy (the orange dashed line marked as $L_{\rm sd}$)
 with typical values $M_{\rm ej}=10^{-4} M_{\odot}$,
$\beta=v_{\rm KN}=0.1c$, $\kappa=1.0~\rm cm^{2}~g^{-1}$, $\tau=10^{4}$ s, and $L_0=1.0 \times 10^{47}~{\rm
erg~s^{-1}}$, and the ejecta$-$CSM interaction (the green dashed line marked as CSM) with $E_{\rm
KN}=10^{49}~{\rm
erg}$, $\kappa_{\rm CSM}=0.33~\rm cm^{2}~g^{-1}$, $z=0.248$, and $M_{\rm CSM}=10 M_{\odot}$. The red solid
line is the sum of the three energy sources.}
 \label{fig:1}
\end{figure}

Figure \ref{fig:1} shows a numerical calculation of the kilonova emission in the $r$ band by
considering the sum of the $r$-process-powered (marked $L_{\rm ra}$), the spin energy from a magnetar
(marked $L_{\rm sd}$), and the ejecta$-$CSM interaction (marked CSM). We find that the most
significant contribution to the kilonova emission from the ejecta$-$CSM interaction is at late times.

\section{Observations of GRB 191019A}
\label{section 3} 
\begin{figure}
 \centering
 \includegraphics [angle=0,scale=0.4] {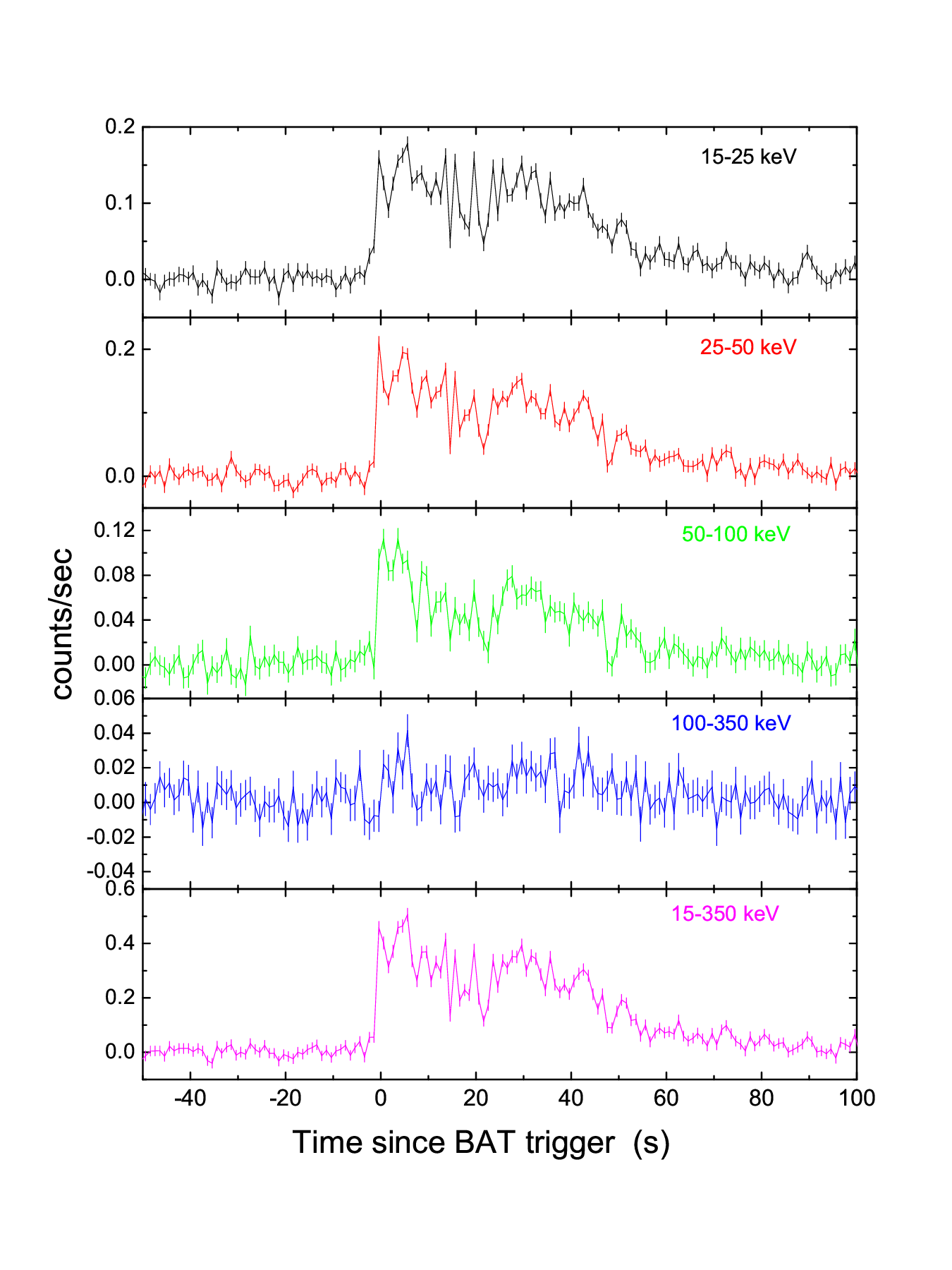}
 \caption{Swift/BAT light curves of GRB 191019A in different energy bands with a 256 ms time bin.}
 \label{fig:2}
\end{figure}

GRB 191019A triggered the BAT at 15:12:33 UT on 2019 October 19
\citep{2019GCN.26031....1S}. The light curve shows a complex structure with many overlapping pulses
(Figure \ref{fig:2}) with duration $T_{\rm 90}= 64\pm4$ s in the energy range $15-350$ keV
\citep{2019GCN.26046....1K}, and thus GRB 191019A is classified as a typical long-duration GRB. The
X-ray Telescope (XRT) began observing the field 3900 $s$ after the BAT trigger
\citep{2019GCN.26045....1D}. Several optical telescopes made follow-up observations of the field,
such as the Ultraviolet and Optical Telescope, the Nordic Optical Telescope (NOT), Gemini-South,
and the Hubble Space Telescope \citep{2023NatAs...7..976L}. Spectroscopy obtained with the NOT and
Gemini-South found a redshift $z=0.248$ within  an old galaxy based on several absorption lines,
and its location was pinpointed within 100 pc projected from the nucleus of its host galaxy
\citep{2023NatAs...7..976L}.

The lack of an associated supernova and the evidence of no star formation in the host galaxy suggest
that a massive star collapse origin of the progenitor of GRB 191019A is disfavored, and that the
progenitor is likely to have arisen either from the merger of compact objects in dynamically dense
stellar clusters or in a gaseous disk around a supermassive BH
\citep{2023NatAs...7..976L}. \citealt{2023ApJ...950L..20L} proposed that GRB 191019A originated
from a compact binary merger, with a prompt emission intrinsic duration of $\sim 1.1$ s, which is
stretched in time by the interaction with a high-density $\sim 10^7-10^8~\rm cm^{-3}$ external
medium.

\section{Possible kilonova emission associated with GRB 191019A }
\label{section 4} If the merger of two compact stars is indeed taking place in GRB 191019A in the
disk of an active galactic nucleus (AGN), the contribution to the kilonova emission from the ejecta$-$CSM
interaction cannot be ignored in such a dense external medium. In this section, we calculate the
emission of the GRB 191019A kilonova by considering both the $r$-process-powered energy and the ejecta$-$CSM
interaction ($L_{\rm ra}+$CSM), or the sum of the $r$-process-powered energy, the spin energy from a
magnetar,
and the ejecta$-$CSM interaction ($L_{\rm ra}+L_{\rm sd}+$CSM).

\begin{figure}
\centering
\includegraphics [angle=0,scale=0.45] {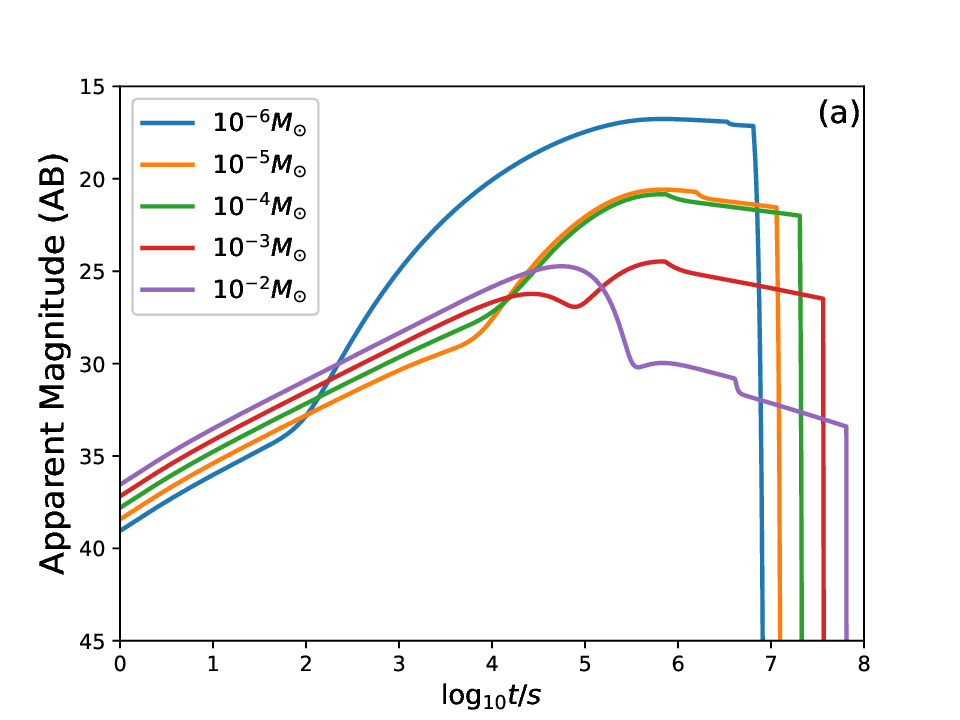}
\includegraphics [angle=0,scale=0.45] {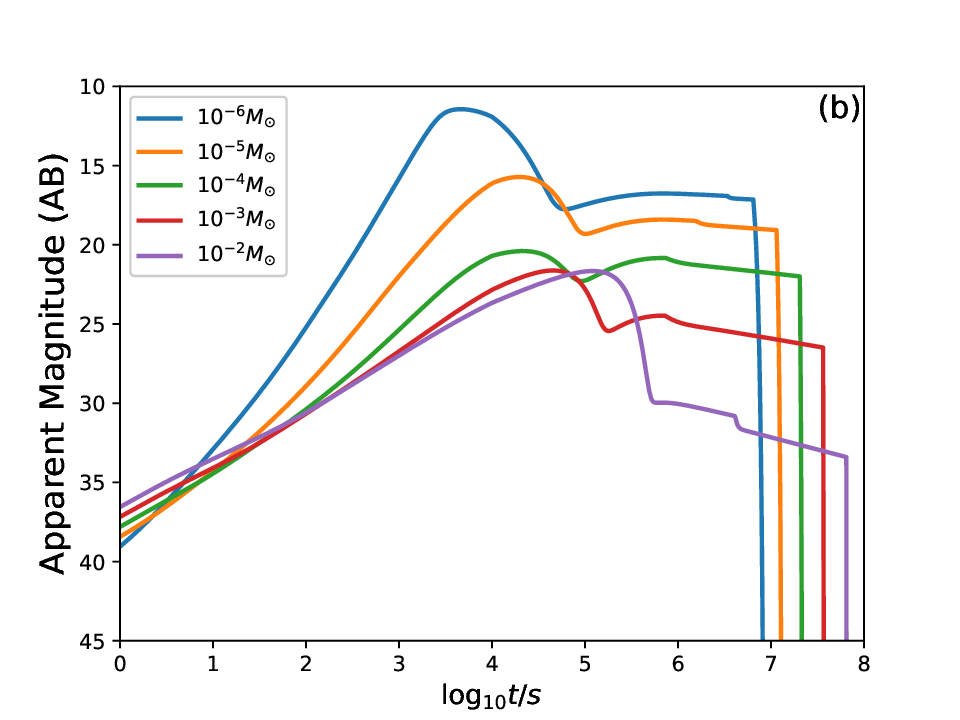}
 \caption{Kilonova $r$-band light curve with variations of $M_{\rm ej}=10^{-2}M_{\odot}$ (purple line),
 $10^{-3}M_{\odot}$ (red line), $10^{-4}M_{\odot}$ (green line), $10^{-5}M_{\odot}$ (orange line), and
 $10^{-6}M_{\odot}$ (blue line). (a) Considering the source energy of $L_{\rm ra}+$CSM. (b) Considering
 the source energy of $L_{\rm ra}+L_{\rm sd}+$CSM. The other parameters are the same as in Figure
 \ref{fig:1}.}
 \label{fig:3}
\end{figure}

\begin{figure}
\centering
\includegraphics [angle=0,scale=0.45] {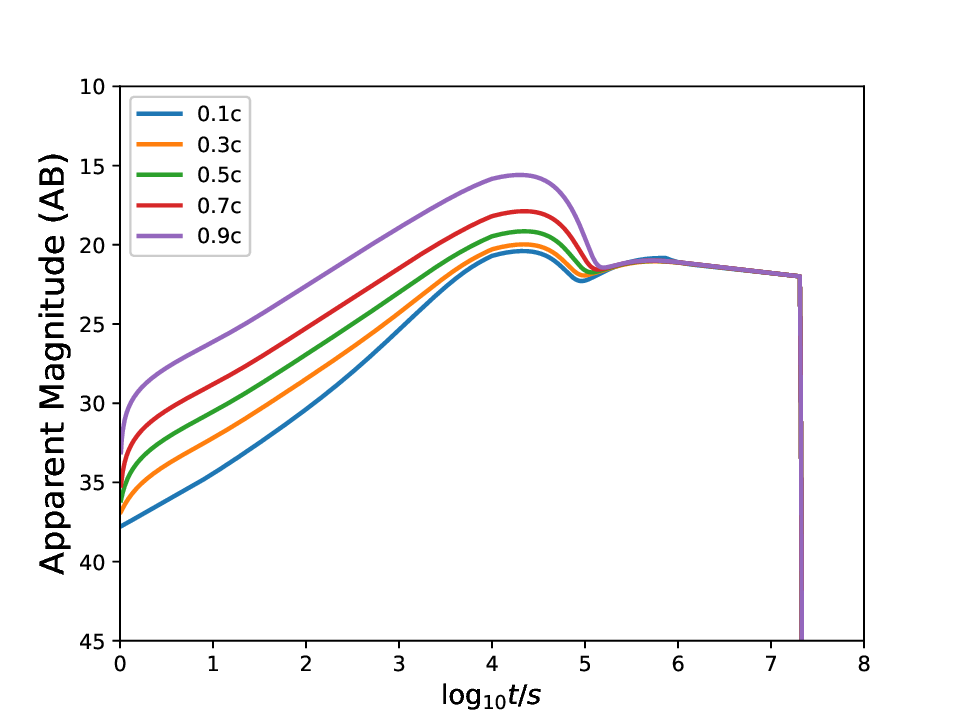}
\includegraphics [angle=0,scale=0.45] {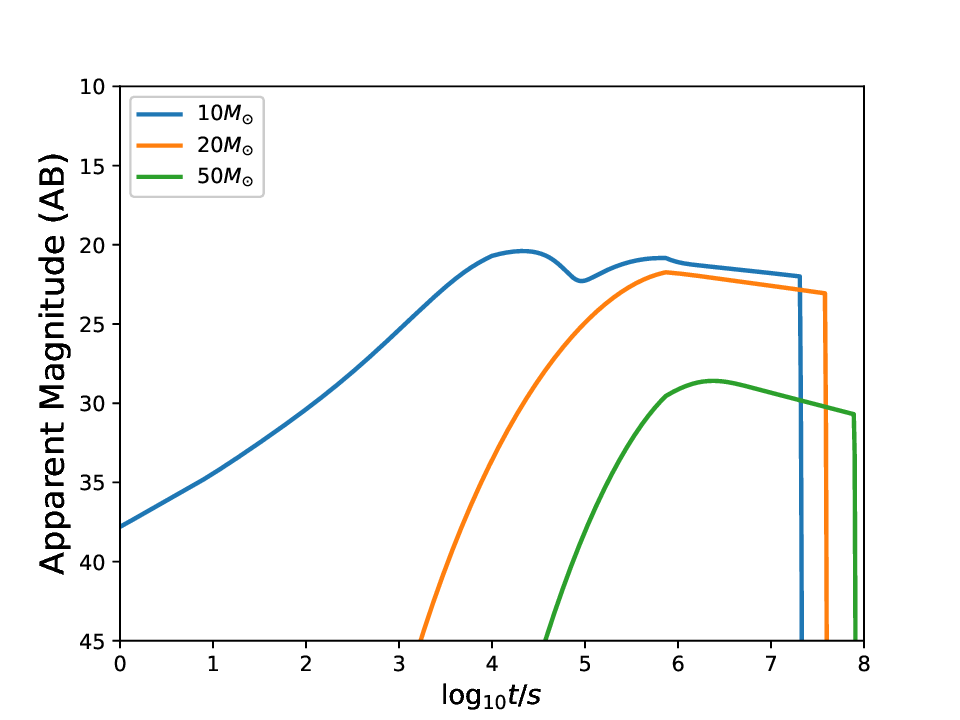}
 \caption{Kilonova $r$-band light curve with variations of ejecta velocity (left) and CSM mass (right).
 The other parameters are the same as in Figure \ref{fig:1}.}
 \label{fig:4}
\end{figure}

\cite{2023ApJ...950L..20L} showed that the number density of the external medium can be as high as
$10^{7}-10^{8}~\rm cm^{3}$. By assuming that the external medium is filled with protons, the volume
density is $\rho=1.7\times10^{-16}\rm g~cm^{-3}$. We calculate the kilonova emission given the
ejecta masses of $M_{\rm ej}=10^{-2}$, $10^{-3}$, $10^{-4}$, $10^{-5}$, and $10^{-6} M_\odot$ by
considering both $L_{\rm ra}+$CSM (the left panel in Figure \ref{fig:3}) and $L_{\rm ra}+L_{\rm sd}+$CSM
(the right panel in Figure \ref{fig:3}). We find that a larger ejecta mass has a stronger $L_{\rm ra}$
radiation, but a smaller ejecta mass corresponds to a brighter kilonova emission contribution from
the ejecta$-$CSM interaction at later time. Moreover, we also show that how the sensitivity of the kilonova
emission is for different values of ejecta velocity and CSM mass (Figure \ref{fig:4}). We find that
the CSM interaction contribution of the kilonova emission is not sensitive to the ejecta velocity, but
depends a little bit on the CSM mass. Here, a large mass of CSM is possible due to external
environmental influence (such as gas drag or migration traps).

One basic question is whether or not we can find evidence for a kilonova emission from an
ejecta$-$CSM interaction from the observations. We collect all the optical data that was observed by
telescopes in \cite{2023NatAs...7..976L} in the $r$, $g$, $z$, $U$, and $B$ bands. We find that the host
galaxy of GRB 191019A is brighter than any afterglow or kilonova emission, except in the $U$ band at
early times. Therefore, it would be difficult to detect any kilonova emission from GRB 191019A. If
this is the case, one interesting question is: what is the parameter distribution for a nondetection
of a kilonova emission from GRB 191019A? Figure \ref{fig:5} shows the possible kilonova emission,
which is dimmer than the host galaxy and cannot be detected within the $L_{\rm ra}+L_{\rm sd}+$CSM
scenario. A nondetection requires that the injected mass is larger than $M_{\rm
ej}=2\times10^{-4}M_{\odot}$ and the injected luminosity and magnetar timescale are less than $L_0=
10^{48}~{\rm erg~s^{-1}}$ and $t_{\rm sd}=10^{3}$ s, respectively.

The other case is that if the host galaxy is not bright enough or the kilonova emission of GRB
191019A is brighter than the host galaxy. Figure \ref{fig:6} shows a possible kilonova emission
which is brighter than the host galaxy and can be detected within the $L_{\rm ra}+L_{\rm sd}+$CSM
scenario. It requires that the injected mass is less than $M_{\rm ej}=2\times10^{-5}M_{\odot}$ and
the injected luminosity and magnetar timescale to be larger than $L_0= 10^{48}~{\rm erg~s^{-1}}$ and
$t_{\rm sd}=10^{3}$ s, respectively. If this is the case, the component of the kilonova emission from
the ejecta$-$CSM interaction can be embodied at the later time.

\section{Conclusion and discussion}
\label{section 5}

The prompt emission light curve of GRB 191019A shows a complex structure with a duration of $T_{\rm
90}= 64\pm4$ s in the energy range $15-350$ keV \citep{2019GCN.26046....1K}, and the GRB is
classified as a typical long-duration GRB. It is quite different from that of typical short GRBs
and short GRBs with EE (e.g. GRBs 060614 and 211211A). However, the lack of an
associated supernova at low redshift $z=0.248$ and the evidence that it lives in a host galaxy with no
star formation suggest that a massive star collapse origin of the progenitor of GRB 191019A is
disfavored, and it likely originated in a compact object merger \citep{2023NatAs...7..976L}.
\cite{2023ApJ...950L..20L} proposed that it may form dynamically in dense stellar clusters or
originate in a gaseous disk around the supermassive BH. If this is the case, the prompt
emission of GRB 191019A can be easily interpreted as having a short intrinsic duration that was
stretched in time by the interaction with a dense external medium and seen as moderately off-axis
\citep{2022ApJ...938L..18L,2023ApJ...950L..20L}.

If GRB 191019A is indeed taking place in such high-density environment after a compact binary
merger (NS-NS or NS-BH) on the AGN's disk, then the contribution to the kilonova emission from an
ejecta$-$CSM interaction cannot be ignored. In this paper, we theoretically calculate the kilonova
emission by considering the contribution of the ejecta$-$CSM interaction in the high-density
environment, the $r$-process-powered energy, as well as the magnetar spin energy. It is found that the
significant
contribution to the kilonova emission from the ejecta$-$CSM interaction occurs at the later time. In
this case, a larger ejecta mass has a stronger $L_{\rm ra}$ radiation, but a smaller ejecta mass
corresponds to a brighter kilonova emission contribution from the ejecta$-$CSM interaction at a later
time. This work is distinct from previous works that have not considered the contribution from the
ejecta$-$CSM interaction to the kilonova emission.

Moreover, we try to apply it to GRB 191019A, but we find that it is difficult to identify the
possible kilonova emission from the observations, due to the contribution of the bright host galaxy.
By assuming that the brightness of the host galaxy is at the upper limit of the kilonova emission
associated with GRB 191019A, the model requires that the injected mass is larger than $M_{\rm
ej}=2\times10^{-4}M_{\odot}$ and the injected luminosity and magnetar timescale are less than $L_0=
10^{48}~{\rm erg~s^{-1}}$ and $t_{\rm sd}=10^{3}$ s, respectively. On the other hand, a detected
associated kilonova emission to GRB 191019A would require the injected mass be less than $M_{\rm
ej}=2\times10^{-5}M_{\odot}$ and the injected luminosity and magnetar timescale to be larger than
$L_0=10^{48}~{\rm erg~s^{-1}}$ and $t_{\rm sd}=10^{3}$ s, respectively. Those values of the parameters
fall into a reasonable range \citep{2015ApJ...812...24F,2017ApJ...836..230C}, and the component of
the kilonova emission due to the ejecta$-$CSM interaction can be embodied at later times if the host
galaxy is not bright enough.

\begin{figure}
\centering
\includegraphics [angle=0,scale=0.3] {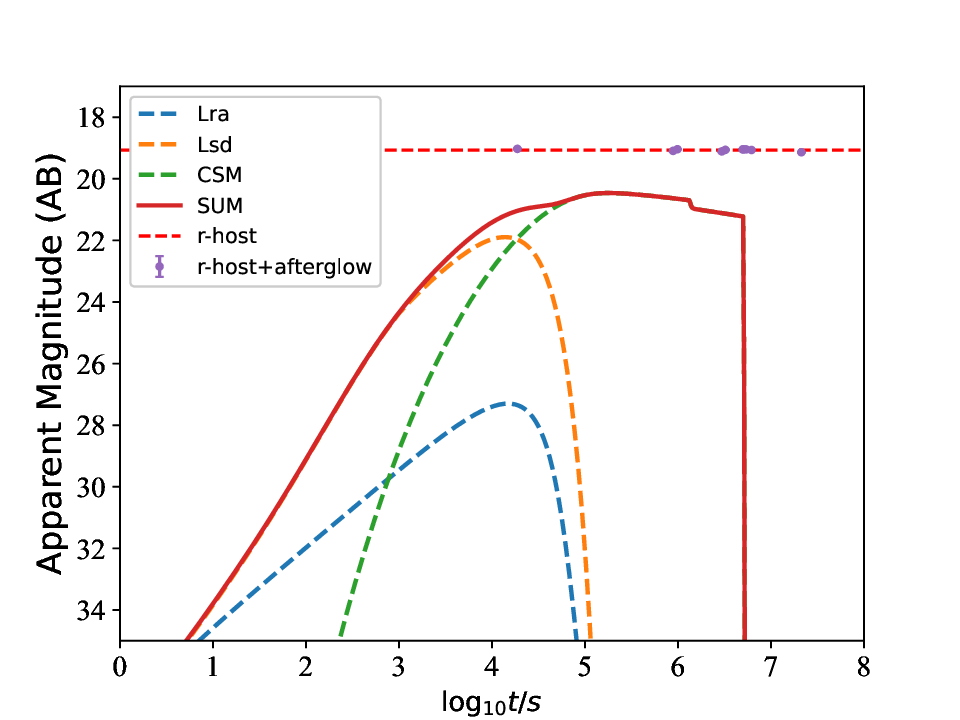}
\includegraphics [angle=0,scale=0.3] {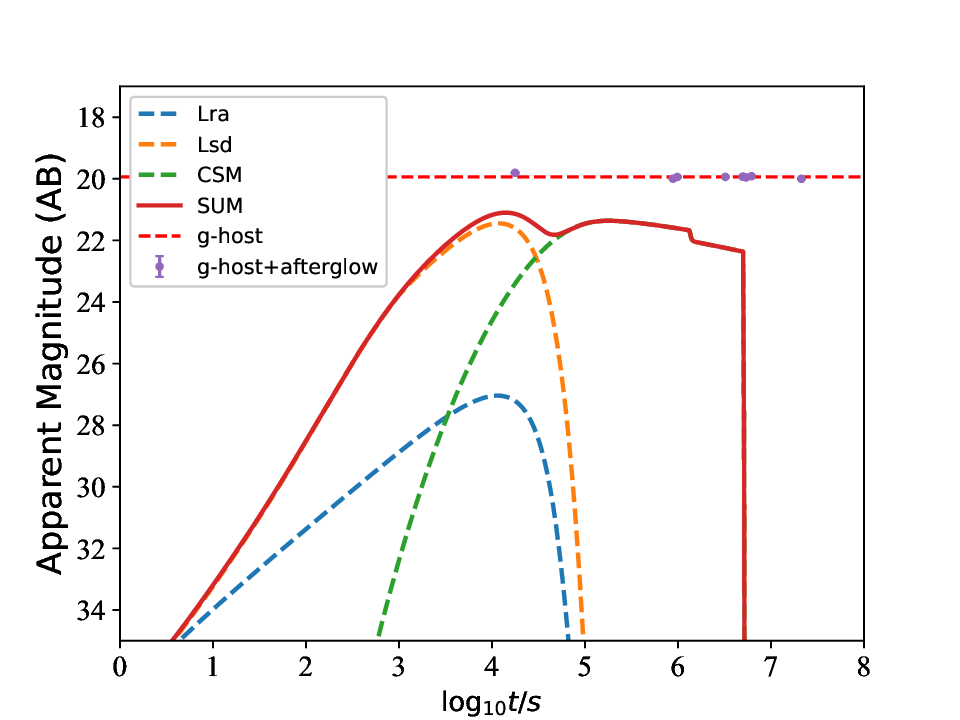}
\includegraphics [angle=0,scale=0.3] {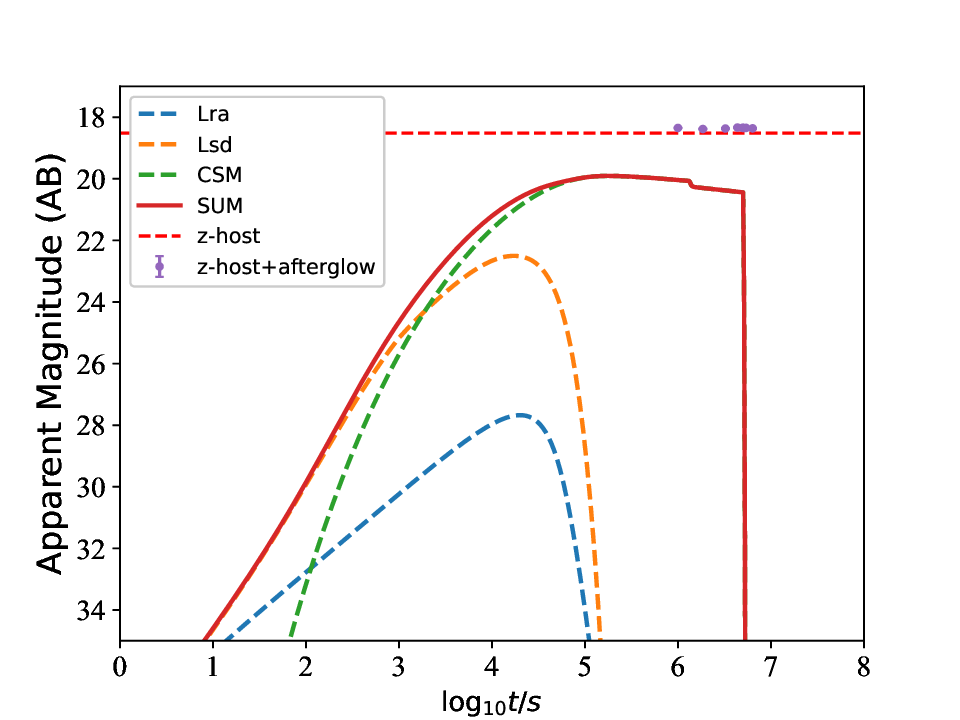}
\includegraphics [angle=0,scale=0.3] {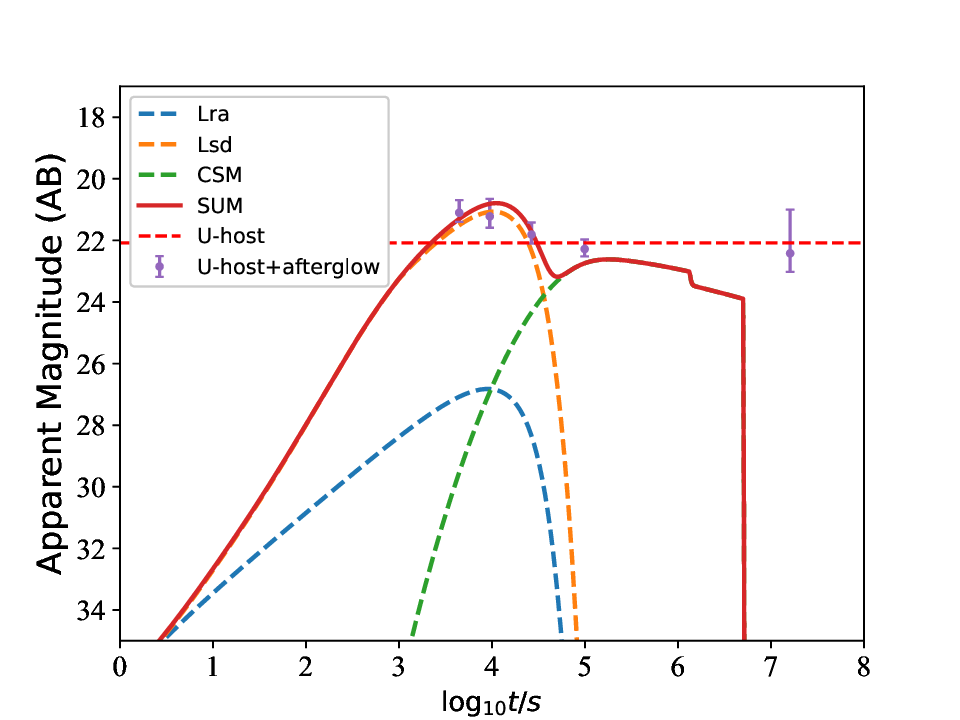}
\includegraphics [angle=0,scale=0.3] {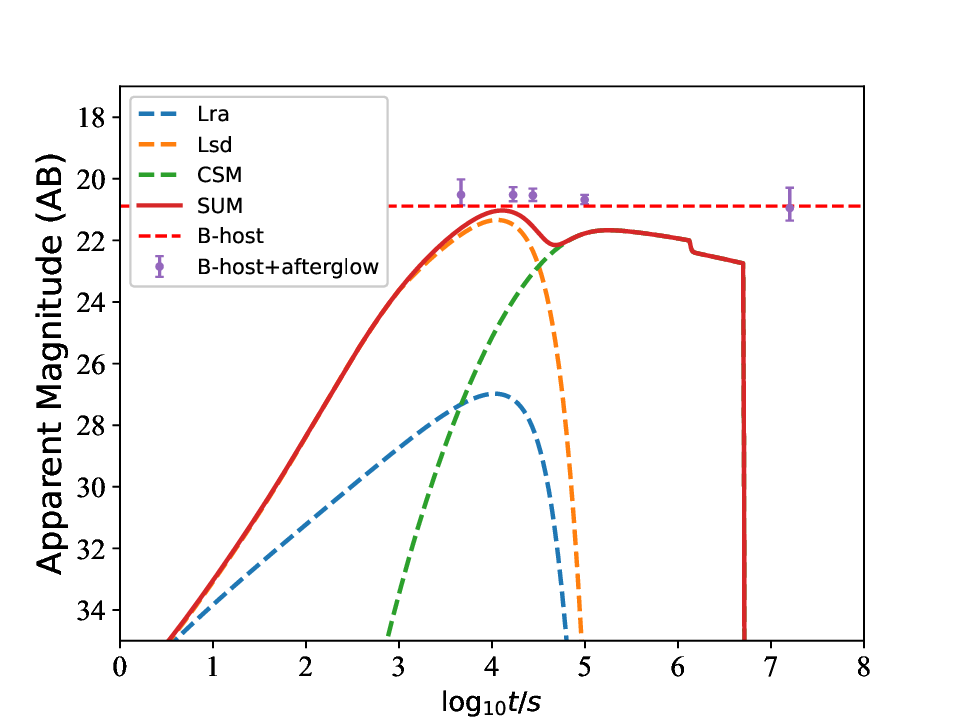}
 \caption{Kilonova light curves in the $r$, $g$, $z$, $U$, and $B$ bands of GRB 191019A, considering
 the three energy sources. A kilonova nondetection requires $M_{\rm ej}=2\times10^{-4}M_{\odot}$,
 $\tau=10^{3}$ s, and $L_0=10^{48}~{\rm erg~s^{-1}}$, $\kappa_{\rm CSM}=0.2~\rm cm^{2}~g^{-1}$, and
 $E_{\rm KN}=3.5\times10^{49}~{\rm erg}$. The red dashed line is the brightness of the host galaxy.
 The purple data points are the observations of host$+$afterglow collected from \cite{2023NatAs...7..976L}.}
 \label{fig:5}
\end{figure}

\begin{figure}
\centering
\includegraphics [angle=0,scale=0.3] {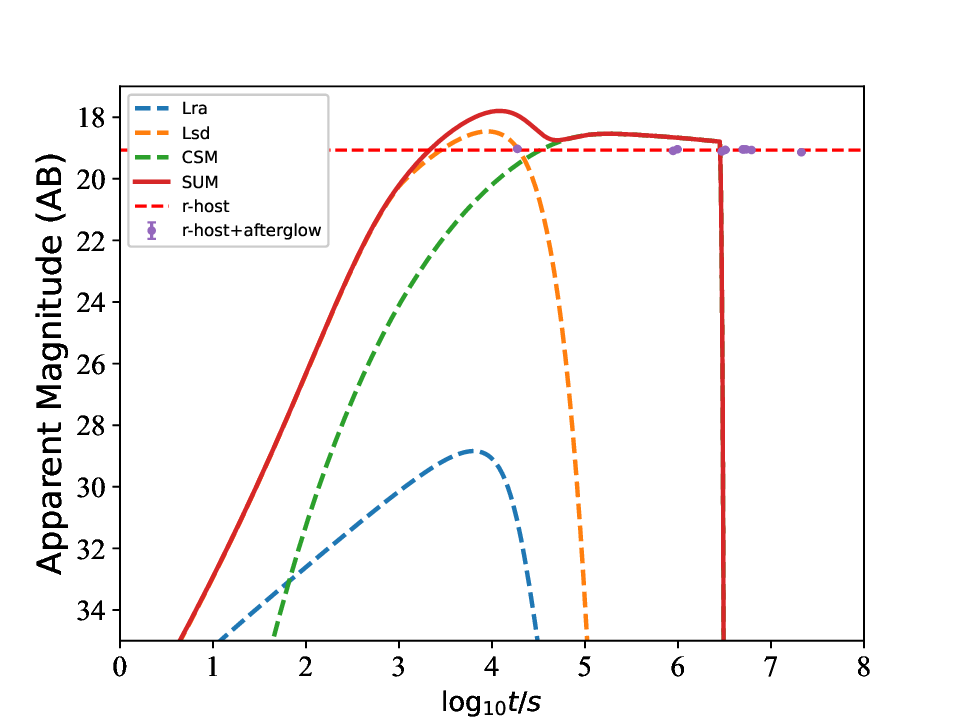}
\includegraphics [angle=0,scale=0.3] {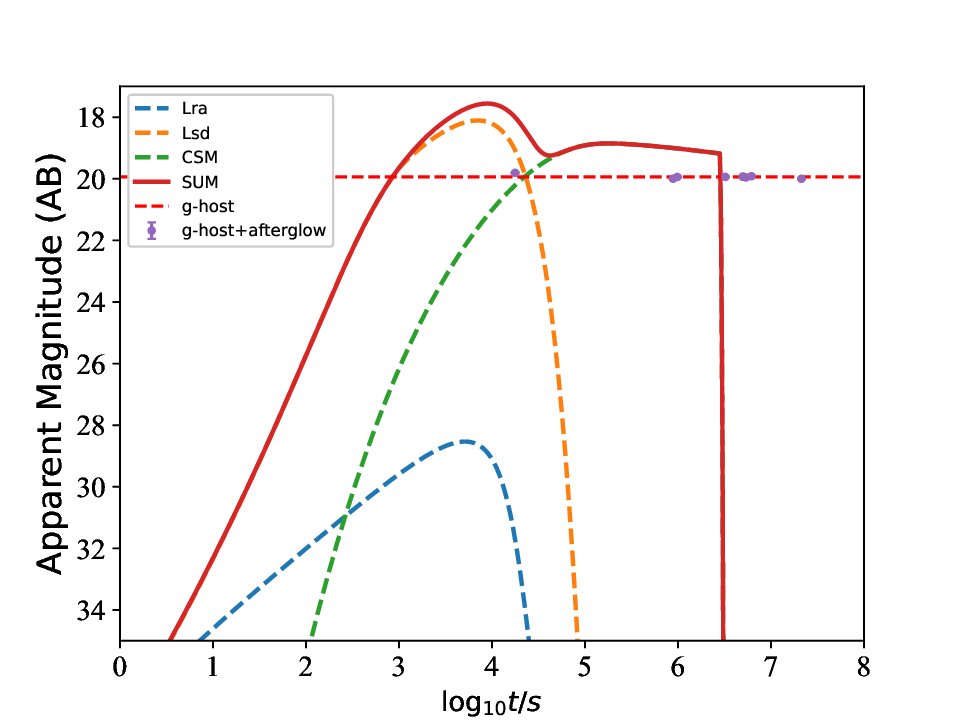}
\includegraphics [angle=0,scale=0.3] {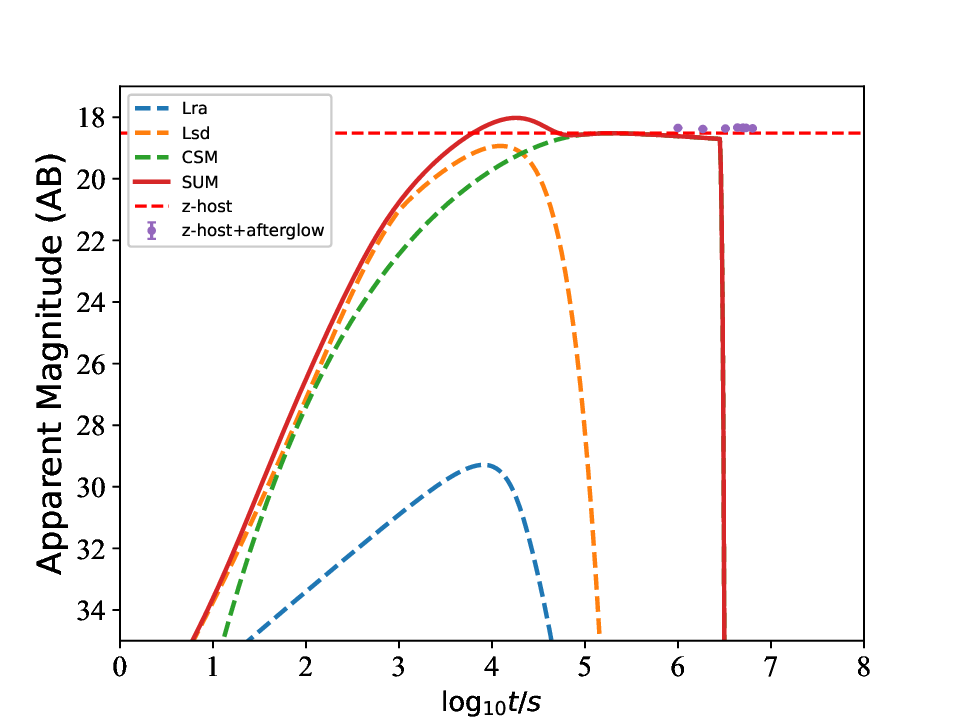}
\includegraphics [angle=0,scale=0.3] {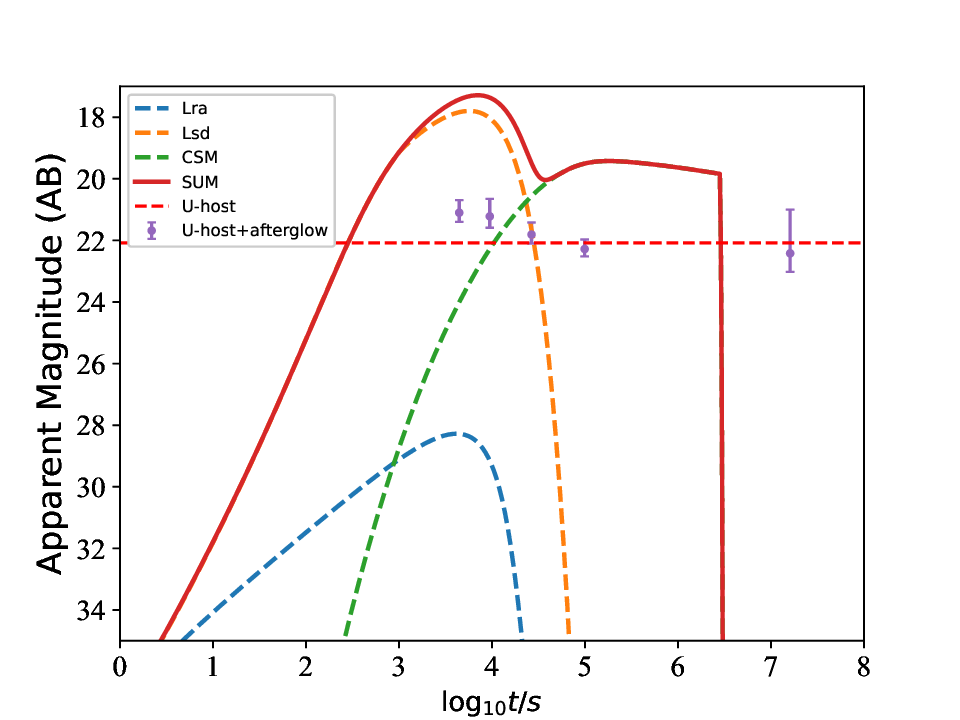}
\includegraphics [angle=0,scale=0.3] {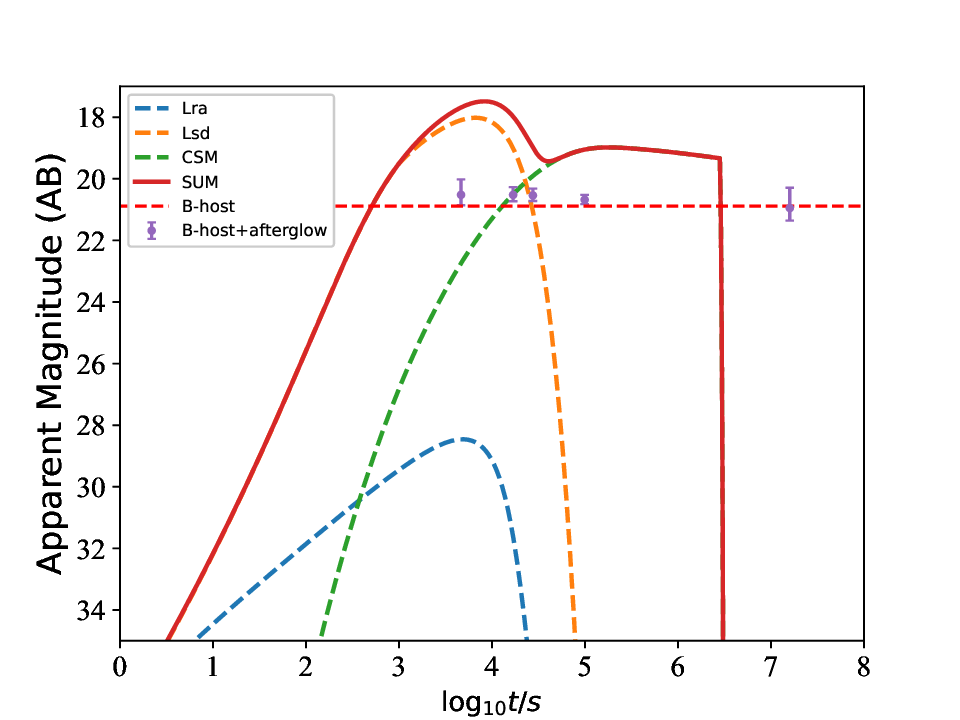}
 \caption{Kilonova light curves that are similar to Figure \ref{fig:5}, but with a detected kilonova
 emission
 with the value of the parameter $M_{\rm ej}=2\times10^{-5}M_{\odot}$.}
 \label{fig:6}
\end{figure}

Our calculations also pose a curious question: is the nondetection of an associated kilonova
emission to GRB 191019A due to a larger optical depth that presents photons from escaping the
system before the medium becomes transparent at the photosphere radius? Based on the parameters we
use in Figure \ref{fig:5} and Figure \ref{fig:6}, we calculate the optical depth of the ejecta$-$CSM
interaction and find that the optical depth is close to or even less than 1. This implies that the
nondetection of any associated kilonova emission from GRB 191019A is not due to a large optical
depth, but is either overwhelmed by the bright contribution from the host galaxy or that the
kilonova emission is itself weak.

In the future, in order to investigate the kilonova emission contribution from the ejecta$-$CSM
interaction associated with a GRB 191019A$-$like event, a brighter kilonova emission (at least
brighter than that of the host galaxy) or an even lower redshift will be required. There are event rate
densities of NS-NS ($R_{\rm NS-NS}\sim f_{\rm AGN} [0.2, 400]\rm~ Gpc^{-3}~ yr^{-1}$) and NS-BH
($R_{\rm NS-BH}\sim f_{\rm AGN} [10, 300]\rm~ Gpc^{-3}~ yr^{-1}$) mergers in the disks of AGNs
\citep{2020MNRAS.498.4088M,2021ApJ...906L..11Z}. They also suggested that only 10\%-20\% of NS-BH
mergers in AGN disks can result in tidal disruptions. If this is the case, the expected a detection
rate is $R_{\rm obs}\sim f_{\rm AGN}f_{\theta} [20, 400]\rm~ yr^{-1}$. Here, $f_{\theta}$ is the
fraction of mergers and $f_{\rm AGN}$ is the fraction of the observed NS-NS or NS-BH mergers from an
AGN disk. We therefore encourage optical follow-up observations of GRB 191019A$-$like events,
especially those that capture electromagnetic signals at later times if the host galaxy is not
bright enough. If possible, an observed GRB 191019A-like event with an associated
GW signal would also provide a good probe for understanding the types of compact
star mergers (e.g., NS-NS or NS-BH).

The James Webb Space Telescope (JWST) has a near-infrared limit of 28. If the contribution of the
host galaxy is very small, current detectors would be able to see such kilonova emission in the
future.

\begin{acknowledgements}
We thank Bing Zhang and Ren Jia for helpful comments. This work is supported by the Guangxi Science
Foundation the National (grant Nos. 2023GXNSFDA026007), the National Natural Science Foundation of China
(grant No. 11922301 and 12133003), and the Program of Bagui Scholars Program (LHJ).
\end{acknowledgements}


\clearpage


\end{document}